# A Comparison between Two Main Academic Literature Collections: Web of Science and Scopus Databases

Arezoo Aghaei Chadegani[1], Hadi Salehi[2], Melor Md Yunus[3], Hadi Farhadi[4], Masood Fooladi[1], Maryam Farhadi[1] & Nader Ale Ebrahim[5]

[1] Department of Accounting, Mobarakeh Branch, Islamic Azad University, Mobarakeh, Isfahan, Iran

[2] Faculty of Literature and Humanities, Najafabad Branch, Islamic Azad University, Najafabad, Isfahan, Iran

[3] Faculty of Education, Universiti Kebangsaan Malaysia (UKM), Malaysia

[4] School of Psychology and Human Development, Faculty of Social Sciences and Humanities, Universiti Kebangsaan Malaysia (UKM), Malaysia

[5] Research Support Unit, Centre of Research Services, Institute of Research Management and Monitoring (IPPP), University of Malaya, Malaysia

Correspondence: Arezoo Aghaei Chadegani, Department of Accounting, Mobarakeh Branch, Islamic Azad University, Mobarakeh, Isfahan, Iran. Tel: 60-17-319-1093. E-mail: arezooaghaie2001@yahoo.com

**Abstract**

Nowadays, the world's scientific community has been publishing an enormous number of papers in different scientific fields. In such environment, it is essential to know which databases are equally efficient and objective for literature searches. It seems that two most extensive databases are Web of Science and Scopus. Besides searching the literature, these two databases used to rank journals in terms of their productivity and the total citations received to indicate the journals impact, prestige or influence. This article attempts to provide a comprehensive comparison of these databases to answer frequent questions which researchers ask, such as: How Web of Science and Scopus are different? In which aspects these two databases are similar? Or, if the researchers are forced to choose one of them, which one should they prefer? For answering these questions, these two databases will be compared based on their qualitative and quantitative characteristics.

## 1. Introduction

Web of Science (WOS) and Scopus are the most widespread databases on different scientific fields which are frequently used for searching the literature (Guz & Rushchitsky, 2009). WOS from Thomson Reuters (ISI) was the only citation database and publication which covers all domains of science for many years. However, Elsevier Science introduced the database Scopus in 2004 and it is rapidly become a good alternative (Vieira & Gomes, 2009). Scopus database is the largest searchable citation and abstract source of searching literature which is continually expanded and updated (Rew, 2009). WOS is challenged by the release of Scopus, an academic literature database which is built on a similar breadth and scale. WOS and Scopus are expensive products and it may not be feasible to obtain both of them. Therefore, by surfacing the new citation database, Scopus, scientific libraries have to decide about which citation database can best meet the requests of the consumers?

The competition between WOS and Scopus databases is intense. This competition has led to improvements in the services offered by them. Recently, various papers have compared the coverage, features and citation analysis capabilities of WOS and Scopus (Bakkalbasi, 2006; Burnham, 2006; LaGuardia, 2005; Deis & Goodman, 2005; Dess, 2006; Li et al., 2010). These comparative studies of WOS and Scopus conclude that these two databases are permanently improving. They also conclude that the significant advantage of choosing one of these two sources depends on the particular subject's area. Some researchers propose undertaking subject's specific analysis to find out which database work best for specific fields or time period (Bar-Ilan et al., 2007; Bakkalbasi et al., 2006; Neuhaus & Daniel, 2008). Based on Lopez-Illescas et al. (2008), prior comparisons of these two databases have not exposed a clear winner. They believe that the advantage of one database over another one depends on what explicitly will be analyzed, the scientific field and also period of analysis. Due to this argument, the existing issue is: which database is superior to use? Which one should you subscribe to? The answer to this question is full of equivocation.

In this article, we will briefly analyze the journal evaluation capabilities of WOS and Scopus databases. In fact, this article concentrates on features and capabilities of each product and these databases will be compared from different aspects such as: 1) Provenance and coverage 2) Searching and analysis of results 3) Citation tracking and citation analysis 4) Forming and costs 5) Impact factors 6) Indexing (h-index) 7) Researcher profile and ID tools. Prior studies which compare these databases will be reviewed in this article.

## 2. Provenance and Coverage

Web of Science (WOS), products from "Thomson Reuters Institute of Scientific Information" (ISI), arises from the Science Citation Index created by Eugene Garfield in 1960s. WOS includes above 10,000 journals and comprises of seven different citation databases including different information collected from journals, conferences, reports, books and book series. WOS citation databases are Social Sciences Citation Index (SSCI), Science Citation Index Expanded (SCI Expanded), Conference Proceedings Citation Index Science (CPCI-S), Arts and Humanities Citation Index (A&HCI) and Conference Proceedings Citation Index-Social Sciences and Humanities (CPCI-SSH). It has also two chemistry databases named Index Chemicus (IC) and Current Chemical Reactions (CCR-Expanded). Since WOS is the oldest citation database, it has strong coverage with citation data and bibliographic data which goes back to 1900 (Boyle & Sherman, 2006). WOS claims it has the most depth and the most quality however, Scopus burst in research domain in 2004 which claims to have the biggest database with the wide range of records.

Scopus, officially named SciVerse Scopus, has introduced by Elsevier in November 2004 to the information market. Scopus is the largest database existing on the market for multidisciplinary scientific literatures (Kuchi, 2004). Scopus covers more than 49 million records including trade publications, open-access journals, and book series. Almost 80% of these records include abstract. It contains 20,500 peer-reviewed journals from 5,000 publishers, together with 1200 Open Access journals, over 600 Trade Publications, 500 Conference Proceedings and 360 book series from all areas of science (Rew, 2009). Scopus offers the new sorting and refining feature for researchers to access above 27 million citations and abstracts going back to 1960s (Boyle & Sherman, 2006). Most of institutions in all over the world such as Latin America, Europe, North America, Australia, Asia and the Middle East believe Scopus have positive influence on their researches. Based on Scopus report more than 50% of its information are from the Middle East, Europe and Africa. Boyle and Sherman (2006) believe that choosing Scopus is due to its quality of outcomes, time savings, ease of use and possible effect on research findings.

Vieira and Gomes (2009) using the set of Portuguese universities compare tow databases. They conclude that Scopus provides 20% more coverage than WOS. Scopus covers broader journal range which is limited to recent articles comparing to WOS. Their results reveal that 2/3 of these records can be found in both databases but 1/3 of the records are only in one database. Vieira and Gomes (2009) suggest the coverage of one journal by Scopus database can have break. It means Scopus has a partial coverage for some journals. Checking the title list with journals indexed in Scopus shows that for different journals, the coverage by Scopus is not complete. The coverage comparison of WOS and Scopus databases provided by free online database comparison tool 'JISC-ADAT' is shown in figure 1.

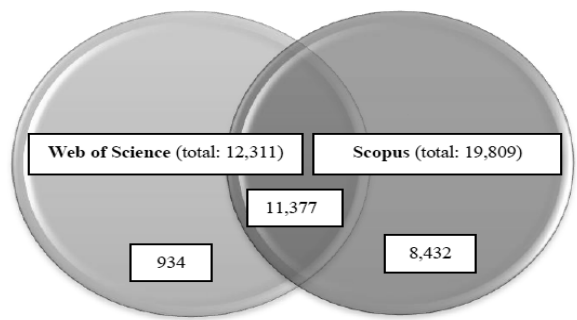

Figure 1. JISC-ADAT coverage comparison of Web of Science and Scopus

## 3. Searching and Analysis of Results

WOS and Scopus are powerful databases which provide different searching and browsing options (Lopez-Illescas et al., 2008). The search options in both databases are the Standard Basic and Advanced. There are different searchable fields and several document types that permit the user to easily narrow their searching. Both databases sort the results by parameters such as; first author, cites, relevance and etc. The Refine Results section in both databases allows the user to quickly limit or exclude results by author, source, year, subject area, document type, institutions, countries, funding agencies and languages. The resulting documents provide a citation, abstract, and references at a minimum. Results may be printed, e-mailed, or exported to a citation manager. The results may also be reorganized according to the needs of the researcher by simply clicking on the headings of each column. Both databases provide users with the ability to set up a profile for personal settings and managing their saved searches and alerts all under one tab. A further user option is the ability to browse individual journals by simply clicking the Journals tab, locate the journal name and select the individual issue. Figure 2 shows the search screen of WOS and Scopus databases.

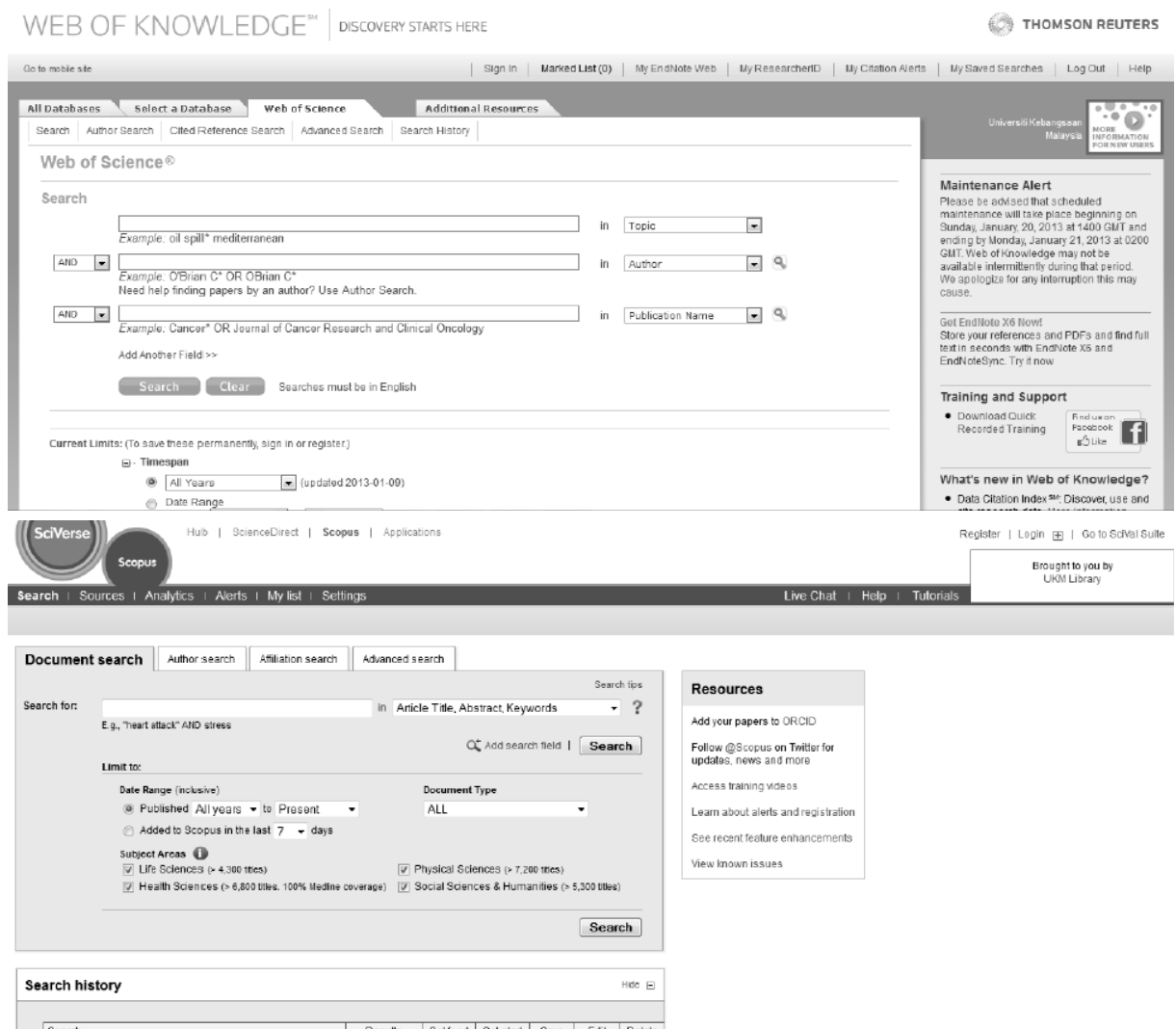

Figure 2. Web of Science and Scopus search screen

The analysis of search results is competitive characteristics of these two databases. Therefore, these databases attempt to improve their features to make searching easier for users. For example, WOS is able to create histograms and rankings according to different parameters. This feature is a stronger tool than that available in Scopus. Moreover, The Scientific Web Plus feature available through WOS enables user to see the search results by clicking on the icon in the search result page. Scientific Web Plus helps users to find scientifically concentrated open web content, author's latest information and research scientists' potential collaborators or competitors, and complements searches. Cited Reference Search (CRS) is another features in WOS for finding previously published and cited articles. WOS also has the "Related Records" feature to detect which authors are citing the same records. The similar feature in Scopus provides the option to discover documents related to the authors or keywords in a specific record.

Another capability of Scopus is journal analyzer which allows users to evaluate journals based on the number of citations, articles published and percentage not cited. One of the weaknesses of Scopus is that it does not have references before 1996. However, one of the important features that introduced by Scopus to online databases is the presentation of faceted indexing. This faceted list of metadata has become a main delight to show in users' training. But, WOS is one step upper than Scopus in representing metadata lists. In fact, search results in WOS are not presented and analyzed by author, year, subject and document types like Scopus, but also by language, institutions and countries in WOS.

## 4. Citation Tracking and Citation Analysis

Citation tracking and analysis have been documented as an important factor for evaluating the influence or importance of specific record for a period of time. Researchers are able to evaluate records using citation data by WOS and Scopus Citation Tracker. In fact, Citation tracking indicates the number of times that a particular work, author or journal have been cited in other works. Citation tracking also allows you to track your own effect and the effect of your institution. Citation tracking offers complete information for other authors and organizations that have similar subject and identifies publications which have similar topics (Quint, 2006).

Citation tracking to journal article is another frequent factor of comparing Scopus and WOS (Bergman, 2012). WOS analyzes the citations by document type, author, funding agency, country, organization's name, language, grant number, publication year and research title. WOS citation reports provide two charts, representing citations in each year and published items in each year. This report also supplies the average citations per item, sum of times cited and the h-index number (Jacso, 2005). Scopus also provides citation information by date of records' citation found about specific article or other records. Similar to WOS, it has a feature to find how many times other authors cited the same topic. Citation overview of Scopus is displayed by option of excluding self-citation with h-index in a graph format (Jacso, 2005). Therefore, both databases allow authors and researchers to discover how many citations an author or an article has received, to find the journals and authors who publish in your area of interest, to explore citations for a specific journal issue, volume or year; to review the citations and work of other authors (Rew, 2009). There are some studies that compare WOS and Scopus databases from the citation aspects and they find different results (Levine & Gil, 2009; Haddow & Genoni, 2010; Bergman, 2012). Figure 3 shows the citation reports in WOS and Scopus.

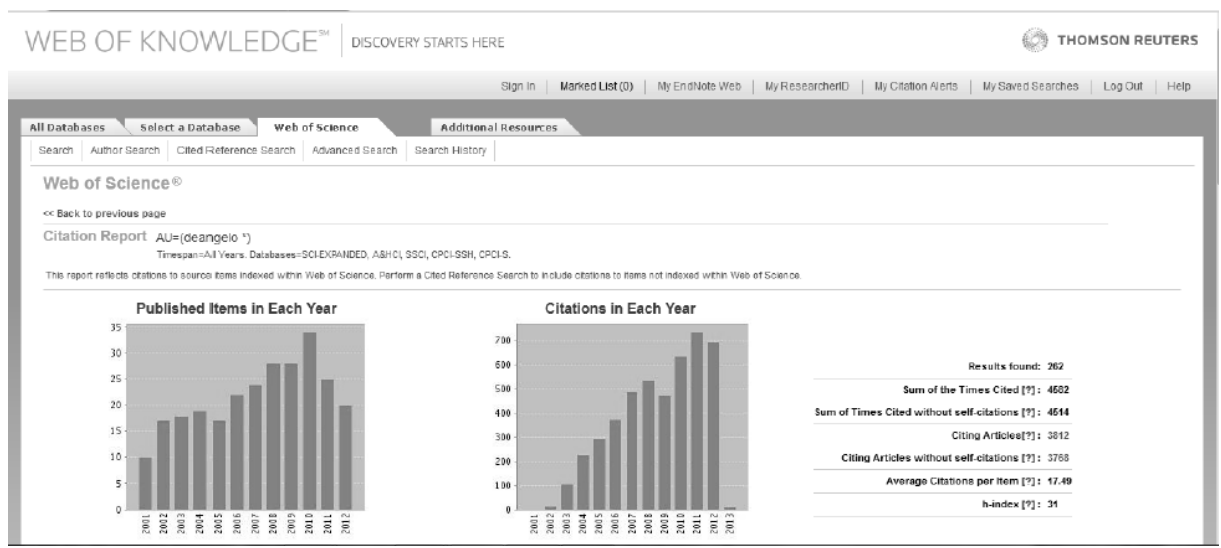

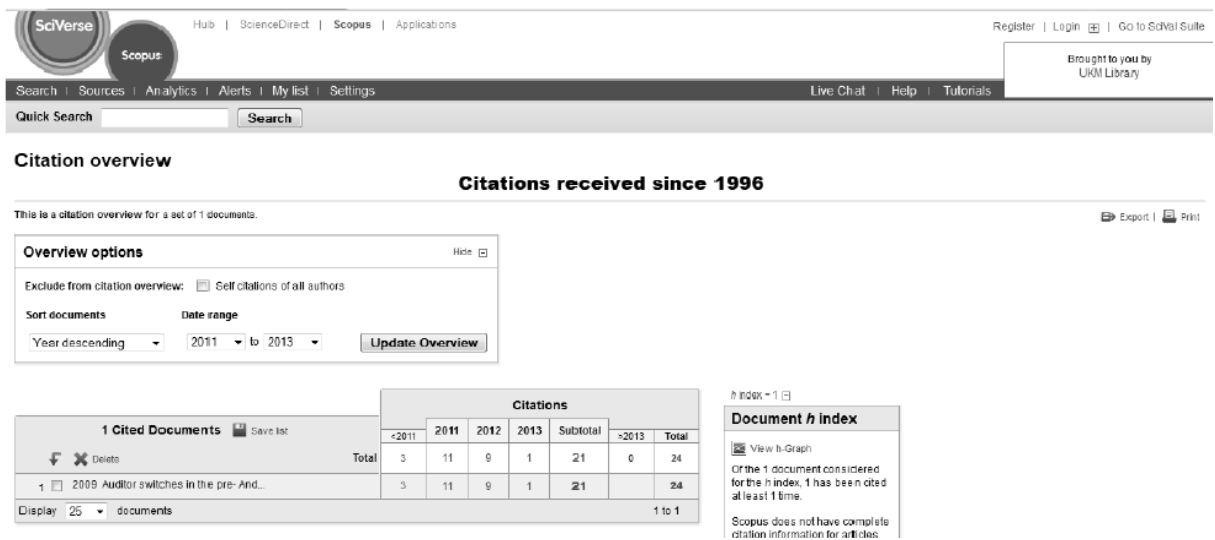

Figure 3. Citation reports in Web of Science and Scopus

Levine Clark and Gil (2009) compare two citation databases for citing references from fifteen business and economic journals. They conclude that Scopus produced slightly higher citation numbers than WOS. Haddow and Genoni (2010) also find that Scopus have more citations counts than WOS and cover more journals. However, Bergman (2012) suggest that for the social welfare journals, Scopus commonly provides higher citation counts than WOS, but the ranking from the highest to the lowest citation count for each database results in similar outcome. Therefore, WOS and Scopus may likely provide correlated results in terms of ranking based on citation counts for other journals. Kulkarni et al. (2009) argue that both WOS and Scopus produce qualitatively and quantitatively diverse citations of articles published in medicine journals. However, Escalona et al. (2010) find that there is a high similarity between WOS and Scopus in other fields such as Chemical Engineering.

Jacso (2005) and Falagas et al. (2008) compare the weaknesses and strengths of WOS and Scopus. They confirm that Scopus covers more journals and its analysis of citations is faster than WOS. However, WOS's citation analysis provides more understandable graphics and is more detailed than Scopus's citation analysis. Meho and Rogers (2008) investigate the differences of Scopus and WOS databases in the citation ranking, citation counting and h-index. They find that no significant differences exist between WOS and Scopus if they are compared only regarding their citation in journals. Levine Clark and Gil (2009) present the result of a comparison of two databases; WOS and Scopus for fifty journals in business and economics. Citations from both databases are analyzed to find out if one database is superior to another one, or whether Scopus could replace WOS. They conclude that researchers may intent to use alternative sources to get a more comprehensive picture of scholarly impact of one article. Bauer and Bakkalbasi (2005) compare the number of citations in WOS and Scopus for American Society Information Science and Technology journal for articles published from 1985 to 2000. They find no difference in citation numbers between WOS and Scopus for articles published in 2000.

## 5. Forming and Costs

Regarding databases forming and shape, Burnham (2006) argues that Scopus is easy to follow even for beginner users comparing WOS. LaGuardia (2005) believes that Scopus draws one and attract researcher in at first look. It's easy to read because fonts and colours provide good contrast and it's less cluttered. Therefore, it is oriented more toward topics searching than WOS.

Pricing is a critical issue as potential buyers pay attention to the cost and benefit outcomes of their decisions. A cost comparison between WOS and Scopus is extremely difficult because information about their cost is closely held by the database producers (Dess, 2006). The results in general show that databases pricing is a complex issue which depends on the size of the organization, discounts which they negotiated and other elements as well. Deis and Goodman (2005) claim that the estimation of WOS costs is about $100,000 per year for large organizations. However, the cost of Scopus database is about 85-95% of the cost of WOS for the same organizations. LaGuardia (2005) argues that WOS refused to provide pricing information but Scopus pricing is set according to annual subscription fee with unlimited usage. Pricing varies based on the size of the

organizations from $20,000 to $120,000 per year. Regarding library budgets, it is highly difficult for any organization to pay for both of these databases. Therefore, the decision about which one to choose is taken by the sort of trade-offs of cost versus performance for each database.

## 6. Impact Factors

The most powerful index which is used by WOS and Scopus databases for ranking journals is the journal's impact factor (IF) (Garfield 2006). IF is calculated according to the citations of all articles published in the journal during two years (Guz & Rushchitsky, 2009). Levine Clark and Gil (2009) define IF as a ratio between citations and recent citable items published. Therefore, the journal's IF is calculated through dividing the numbers of this year citation to the source item which is published in that journal during the prior two years.

Researchers can obtain journals' IF by using two online databases. WOS, Thomson Reuters's product, annually publishes and analyzes indexed journals' IF. Scopus, Elsevier publisher's product, also records the information about journal citations, but does not report indexed journals' IF, which could be obtained through manual calculation of journal citations or other tools (Chou, 2012). Nowadays, it is an honour for each journal to indicate the IF determined by the WOS. However, some scholars have expressed concern that WOS's IF does not accurately reflect the true influence of social work journals within the discipline partially due to the fact that many critical journals are left out of the ranking and calculations (Bergman, 2012).

Lopez Illescas et al. (2008) find that for journals indexed in WOS and Scopus databases, those in WOS indicate higher IFs. On the contrary, those that are only covered by Scopus indicate lower IFs than if they are in both databases. Furthermore, they find that the differences between two databases regarding citation are much lower than the differences regarding coverage. Abrizah et al. (2012) compare the ranking, coverage, IF and subject categorization of Information Science journals based on data from WOS and Scopus. These comparisons are made according to factor scores reported in 2010 Journal Citation Reports. They use the Library of Congress Classification System to compare IF and subject categorization. They find that there is high degree of similarity in rank normalized IF of titles in both WOS and Scopus databases. Pislyakov (2009) reports that the IF ranking of indexed journals between WOS and Scopus is partially different. However, Gary and Hodkinson (2008) conclude that there is no significant difference between WOS and Scopus on journals' IF ranking but these studies only focus on business and science field. Meho and Sugimoto (2009) investigate differences between WOS and Scopus to evaluate the scholarly impact of fields which focus on frequently research domains and institutions, citing journals, conference proceedings, and all citing countries. Their results indicate that when they assess the smaller citing entities such as journals, institutions and conference proceedings, both databases produce significantly different results. However, when they assess larger citing entities such as research domains and countries, they produce similar scholarly impact.

## 7. Indexing (H-Index)

WOS database is a more scholarly source than Scopus database because of more indexing (Fingerman, 2006). Some indices are proposed by WOS and Scopus databases such as h-index, g-index, impact factor, the Eigen factor metric for journal ranking, source normalized impact per paper and relative measure index (Hirsch, 2005; Egghe, 2006; Bergstrom, 2007; Moed, 2010; Raj & Zainab, 2012). The h-index is a well known metric for assessing the researcher's scientific effects (Raj & Zainab, 2012). For measuring h-index, the publication records of an author, the number of papers published during the selected number of years and the number of citations for each paper are considered (Moed, 2010). Glanzel (2006) believes that the advantage of h-index is that it combines the assessment of both the number of papers (quantity) and the impact or citations to these papers (quality). The h-index is automatically computed in both databases for every author and collections of articles which are selected by the user.

Jacso (2011) verifies h-index, which is used by researchers to measure research productivity, for the Scopus database. The results reveal that 18.7 million records in Scopus have one or more cited references, which represents %42 of the whole database content. The cited references enhanced records ratio is rising from 1996 to 2009. Moreover, for other records of 23,455,354 published after 1995, the h-index is 1,339. Therefore, the total number of citations should be at least 1,792,921. For the whole Scopus database of 44.5 million records the h-index is 1,757 (Jacso, 2011). Bar-Ilan (2008) compares the h-index of a list of highly cited authors based on citations counts recovered from the WOS and Scopus. The results show that there is no difference when citation tool is used to calculate the h-index of scientists because the results of both databases are very close to each other. Moreover, it seems to be disciplinary differences in the coverage of the databases. The differences in citation numbers put science policy makers and promotion committees into the trouble.

## 8. Researcher Profile and ID Tools

Both WOS and Scopus databases make available an author search feature by distinguishing the author's affiliation, subject and journals to classify variations for the same author. They also have some different features for researchers to upload their profile information. The Distinct Author Set is a discovery tool in WOS database which shows several papers which written by one author. In fact, WOS analyzes the citation data, like journals and subject areas for developing the distinct author set. However, in Scopus, there is a feature called The Author Identifier. The user can enter the author's name and will have an exact match of whole name of author, as well as how his name appears in journal articles and other records. The Scopus Author Identifier (AI) matches author names according to their affiliation, source title, subject area, address and co-authors. Consequently, the Scopus AI helps researchers to get accurate and comprehensive information as quickly as possible without wasting time on searching through long list of author's names.

Moreover, the research ID feature in WOS allows users to upload their profile information (Vieira & Gomes, 2009). Researcher ID provides a solution to author ambiguity problem in the scholarly research community. Each member is assigned with unique identifier to enable researcher to manage their publication lists, track their times cited counts and h-index, identify potential collaborators and avoid author misidentification. Additionally, your Researcher ID information integrates with the Web of Knowledge and is ORCID compliant, allowing you to claim and showcase your publications from a single one account. Figure 4 shows Researcher ID feature in WOS.

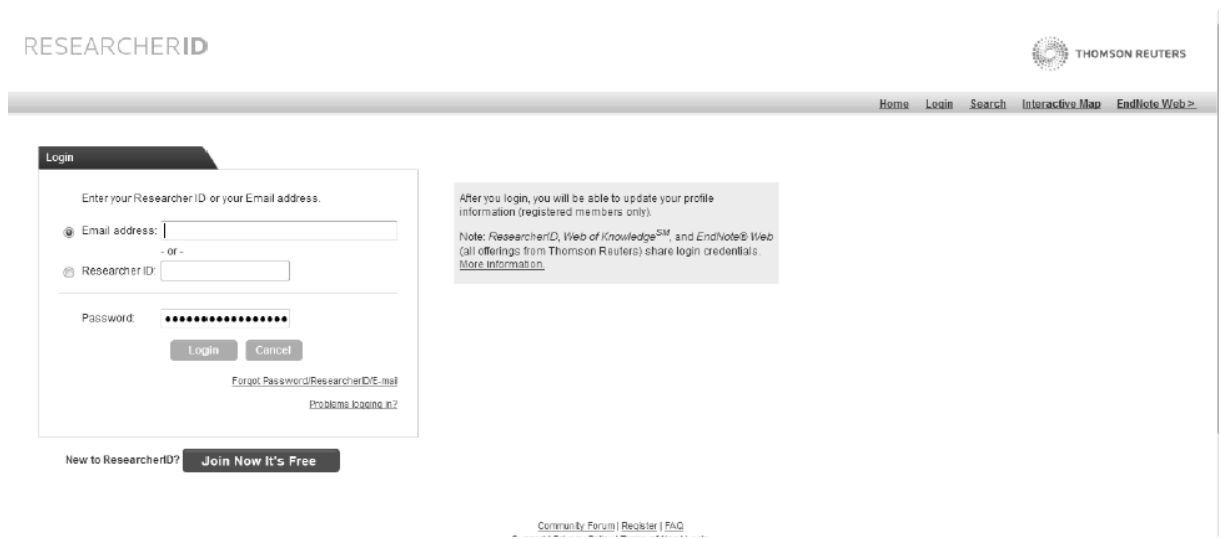

Figure 4. Researcher ID in Web of Science

## 9. Conclusion

Until 2004, the Web of Science (WOS) was the only international and multidisciplinary database available to obtain the literature of technology, science, medicine and other fields. However, Elsevier introduced Scopus which is become a good replacement (Vieira & Gomes, 2009). The information provide by these databases specify the active journals in covering current and relevant research as well as prominent in shaping potential research fields. The intense competition between these databases motivated researchers to compare them to identify their similarities and differences. A numerous researches compare these databases from different aspects. In this article, WOS and Scopus databases are compared based on qualitative and quantitative characteristics such as provenance, citations, searching and special features by reviewing prior studies.

The comparison of WOS and Scopus discovers that WOS has strong coverage which goes back to 1990 and most of its journals written in English. However, Scopus covers a superior number of journals but with lower impact and limited to recent articles. Both databases allow searching and sorting the results by expected parameters such as first author, citation, institution and etc. regarding impact factor and h-index, different results obtained from prior studies. Although there is a high association between both databases, researchers interested to know why authors prefer one database over the other one. For further studies, it is suggested to investigate the perceptions of authors and researchers on both databases to find the reasons which make them to use one database more than the other one. It could be helped databases to improve their features to provide better facilities.